\documentstyle[11pt,newpasp,twoside,epsf]{article}
\def\edcomment#1{\iffalse\marginpar{\raggedright\sl#1\/}\else\relax\fi}
\reversemarginpar

\begin{document}
\title{The Effect of Metallicity on the Cepheid Period-Luminosity Relation}
\author{Martino Romaniello, Francesca Primas, Marta Mottini}
  \affil{European Southern Observatory, Karl-Schwarzschild-Stra{\ss}e 2,
   D-85748 Garching bei M{\"u}nchen, Germany}
\author{Martin Groenewegen}
  \affil{Instituut voor Sterrenkunde, Celestijnenlaan 200B, B-3001
  Leuven, Belgium}

\begin{abstract}
We present preliminary results of an observational campaign devoted at
establishing the influence of chemical composition on the Cepheid
Period-Luminosity relation. The data are in good agreement with
theoretical predictions based on non-linear convective models,
suggesting a fairly strong dependence of the Period-Luminosity
relation on metallicity in the sense of more metal rich stars being
intrinsically fainter than otherwise expected. Our data indicate that
the error on the inferred distance can be as large as 10\% if the role
of metallicity is neglected.
\end{abstract}

\section{Introduction}
Ever since the work of Edwin Hubble, the Cepheid Period-Luminosity
($PL$) relation is a fundamental tool in determining Galactic and
extragalactic distances. In spite of its paramount importance, to this
day we still lack firm theoretical and empirical assessment on whether
or not chemical composition has any significant influence on the
pulsational properties of Cepheids.

The debate on metallicity effects on the Period-Luminosity relation is
an old one (see, for example, the reviews by Madore \& Freedman 1991
and, more recently, Feast 2003), but the matter is far from being
settled. As an illustration, different prescriptions on how to account
for the metallicity effect lead to differences as large as 10\% in the
the distance to the Large Magellanic Cloud (e.g. Groenewegen \&
Oudmaijer 2000), which directly translates into a 10\% uncertainty on
$H_0$. To make things worse, this effect is \emph{systematic} and only
a careful calibration of the dependence of the $PL$ relation on
metallicity can provide firmer ground on which to build the
astronomical distance ladder. Regrettably, the current understanding
of the subject is poor.

Theoretical pulsational models by different groups lead to markedly
different results. On the one side computations based on \emph{linear
models} (e.g. Chiosi et al 1992, Sandage et al 1999, Alibert et al
1999, Baraffe \& Alibert 2001) suggest a mild dependence of the $PL$
relation on chemical composition (less than 0.1 mag at $\log(P)=1$ for
a change in metallicity between $Z=0.004$ and $Z=0.02$).  This result
is challenged by the outcome of the \emph{non-linear convective
models} (e.g. Bono et al 1999, Caputo et al 2000) which find that both
the slope and the zeropoint of the Period-Luminosity relation depend
significantly on the adopted chemical composition. Again for
$\log(P)=1$ and the same variation in metallicity as above, they
predict a change as large as 0.4 magnitudes in $V$, 0.3 magnitudes in
$I$ and $0.2$ magnitudes in $K$. Moreover, the change is such that
metal-rich Cepheids are {\it fainter} than metal-poor ones, again at
variance with the results of Alibert et al (1999) and Baraffe \&
Alibert (2001).  Finally, recent calculations by Fiorentino et al
(2002) based on non-linear models indicate that the $PL$ relation also
depends on the helium content of the stars.

From an observational standpoint, the subject was approached in
essentially two ways: a direct measurement of the iron content in
nearby, \emph{i.e.}  bright, Cepheids, or by measuring a secondary
metallicity indicator in external galaxies known to contain Cepheids,
under the assumption that they would have the same chemical
composition.  Regrettably, \emph{direct metallicity determinations}
for individual Cepheids exist only for a handful of stars. The most
accurate studies are by Fry \& Carney (1997, [Fe/H] and [$\alpha$/H]
for 23 Galactic Cepheids) and Andrievsky et al (2002, complete
abundance analysis of 77 Galactic Cepheids; see also Luck et al
2003). The former find a spread in [Fe/H] of 0.4 dex, which they argue
is real, the latter focus on investigating Galactic abundance
gradients ($-0.05$~dex/kpc). Luck et al (1998) have measured the iron
abundance of 16 Cepheids in the Large and Small Magellanic Cloud (LMC
and SMC, respectively) with the echelle spectrograph at the CTIO 4
meter telescope. The relatively small size of the telescope forced
them to choose only stars at the bright end of the Period-Luminosity
relation, \emph{i.e.} long periods, whereas the entire range needs to
be sample in order to reach meaningful conclusions on the $PL$
relation (e.g. Kennicut et al 1998, Caputo et al 2000).  As for the
\emph{indirect measurements} in external galaxies (e.g. Kochanek et al
1997, Kennicutt et al 1998), which probe a broader range in
metallicity than the one available in the Galaxy, they find that
metal-rich Cepheids are {\it brighter} in $V$ and $I$ than metal-poor
ones by $0.24$ to $0.40$ magnitudes per dex in metallicity. These
results have been widely used in the literature, including in the
final paper of the {\sc hst} Key Project on $H_0$ (Freedman et al
2001).

\section{The Data Sample}
With the scenario described above in mind, we have collected high
quality spectra of an unprecedented dataset of 40 Galactic, 22 LMC and
14 SMC Cepheids with the aim of measuring metallicity directly on a
statistically significant sample of stars covering about one order of
magnitude in chemical composition. The Galactic stars were observed
with FEROS at the ESO 1.5m telescope on La Silla at a spectral
resolution of 48,000 with a signal-to-noise ratio of about 70 to
150. As for the spectra of the Magellanic Cloud objects, we used the
UVES spectrograph on the Kueyen telescope on Cerro Paranal, yielding a
resolution of 40,000 and signal-to-noise ratio of about 50 to 70.

Deriving metallicities for Cepheids is by no means a trivial task, but
we have now carefully analyzed and derived [Fe/H] for an initial
subsample of 14 Cepheids in the LMC, 12 in the SMC and 6 in the
Galaxy. The data reduction procedures and analysis are described in
detail in Mottini et al (2003).

\section{Results: dependence of the PL relation on metallicity}
The main result from the subset analyzed so far is summarized in
Figure~1.  There, we plot the V-band residuals from the standard
Madore \& Freedman (1991) $PL$ relation as a function of the iron
abundance derived from FEROS and UVES spectra. A positive magnitude
difference here means that the star is fainter than the mean $PL$
relation. For consistency with the adopted $PL$ relation, the LMC is
assumed to have a distance modulus of 18.50. The SMC is considered
0.44~mag more distant (e.g. Cioni et al 2000) and the distances to
Galactic stars were taken from Fouqu\'e et al (2003), Barnes et al
(2003) and Tammann et al (2003).  Photometry and reddening are taken
from Laney \& Stobie (1994).

\begin{figure}[!ht]
  \plotfiddle{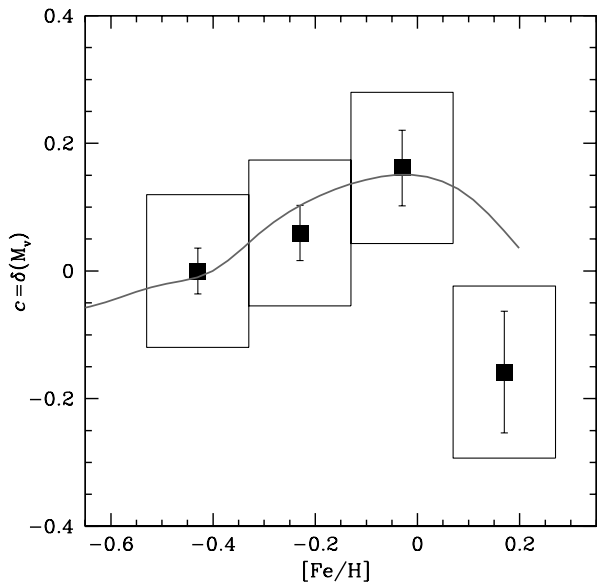}{8cm}{0}{40}{40}{-140}{-40}
  \vspace{-1cm}
  \caption{$V$-band residuals compared to the Madore \& Freedman
  (1991) $PL$ relation vs the iron content we have measured from our
  UVES and FEROS spectra (14 stars in the LMC, 12 in the SMC and 6 in
  the Galaxy). Filled squares represent the median value in each
  metallicity bin with its associated errorbar and the rms in each bin
  is displayed as boxes. The solid line shows the predictions by
  Fiorentino et al (2002).}
\end{figure}

The data are binned in metallicity to reflect the typical
observational error on our determination ($\pm0.1$~dex). The
Magellanic Cloud Cepheids populate the bins at lower metallicity, the
higher [Fe/H] one containing only Galactic stars. The median magnitude
residual in each bin is plotted as a square and the rms and error on
the median are shown as boxes and errorbars, respectively.  The
magnitude difference plotted on the ordinates can be directly compared
to the \emph{``c''} quantity computed by Fiorentino et al (2002). This
is the correction to be applied to the distance modulus derived using
a universal, metallicity independent $PL$ relation to take into
account the influence of metallicity as derived from their non-linear
pulsation models.

The agreement between our data and the predictions by Fiorentino et al
(2002), indicating a strong dependence of the $PL$ relation on
metallicity, is tantalizing, even though the errorbars are still
somewhat large. As it can be seen, the error on the inferred distance
can be as large as 10\% if the effect of metallicity is neglected.


\begin{references}
\reference {Alibert, Y. et al 1999, \aap 344, 551}

\reference {Andrievsky, S.M. et al 2002, \aap, 381, 32}

\reference {Baraffe, I. \& Alibert, Y. 2001, \aap, 371, 592}

\reference {Barnes, T.G. et al 2003, \apj, 592, 539}

\reference {Bono, G. et al 1999, \apj, 512, 711}

\reference {Caputo, F. et al 2000, \aap, 359, 1059}

\reference {Chiosi, C. et al 1992, \apj, 387, 320}

\reference {Cioni, M.-R. et al 2000, \aap, 359, 614}

\reference {Feast, M. 2003, in Stellar Candles for the Extragalactic
  Distance Scale, Lect.Notes Phys., 635, 45}

\reference {Fiorentino, G. et al 2002, \apj, 576, 402}

\reference {Fouqu\'e, P. et al 2003 in Stellar Candles for the Extragalactic
            Distance Scale, Lect.Notes Phys., 635, 21}

\reference {Fry, A.M. \& Carney, B.W. 1997, \aj, 113, 1973}

\reference {Groenewegen, M. \& Oudmaijer, R.D. 2000, \aap, 356, 849}

\reference {Kennicutt, R.C. et al 1998, \apj, 498, 181}

\reference {Kochanek, C.S. 1997, \apj, 491, 13}

\reference {Laney, C.D. \& Stobie, R.S. 1994, \mnras, 266, 441}

\reference {Luck, R.E. et al 1998, \aj, 115, 605}

\reference {Luck, R.E. et al 2003, \aap, 401, 939}

\reference {Madore, B. \& Freedman, W. 1991, \pasp, 103, 933}

\reference {Mottini, M. et al 2003, this conference}

\reference {Sandage, A. et al 1999, \apj, 522, 250}

\reference {Tammann, G. et al 2003, \aap, 404, 423}

\end{references}
\end{document}